\pdfoutput=1

\documentclass[twocolumn, switch]{article} 

\usepackage{preprint}

\newcommand{\btheta}{ \mbox{\boldmath $\theta$}}

\newcommand{\bxi}{ \mbox{\boldmath $\xi$}}

\newcommand{\bXi}{ \mbox{\boldmath $\Xi$}}

\newcommand{\bx}{ \mbox{\bf x}}

\newcommand{\bs}{ \mbox{\bf s}}

\newcommand{\bI}{ \mbox{\bf I}}

\newcommand{\indep}{\stackrel{indep}{\sim}}

\newcommand{\argmin}{{\mathop{\rm arg\, min}}}

\newcommand{\beq}{ \begin{equation}}
\newcommand{\eeq}{ \end{equation}}
\newcommand{\beqn}{ \begin{eqnarray}}
\newcommand{\eeqn}{ \end{eqnarray}}

\usepackage{amsmath, amsthm, amssymb, amsfonts}

\usepackage[numbers,square]{natbib}
\usepackage[utf8]{inputenc}	
\usepackage[T1]{fontenc}	
\usepackage{xcolor}		
\usepackage[colorlinks = true,
            linkcolor = purple,
            urlcolor  = blue,
            citecolor = cyan,
            anchorcolor = black]{hyperref}	
\usepackage{booktabs} 		
\usepackage{nicefrac}		
\usepackage{microtype}		
\usepackage{lineno}		
\usepackage{float}			
\usepackage{arydshln}
\usepackage{subcaption}

\usepackage{lipsum}		

\usepackage{newfloat}
\DeclareFloatingEnvironment[name={Supplementary Figure}]{suppfigure}
\usepackage{sidecap}
\sidecaptionvpos{figure}{c}

\usepackage{titlesec}
\titlespacing\section{0pt}{12pt plus 3pt minus 3pt}{1pt plus 1pt minus 1pt}
\titlespacing\subsection{0pt}{10pt plus 3pt minus 3pt}{1pt plus 1pt minus 1pt}
\titlespacing\subsubsection{0pt}{8pt plus 3pt minus 3pt}{1pt plus 1pt minus 1pt}

\usepackage{tikz,xcolor,hyperref}

\definecolor{lime}{HTML}{A6CE39}
\DeclareRobustCommand{\orcidicon}{
	\begin{tikzpicture}
	\draw[lime, fill=lime] (0,0)
	circle [radius=0.16]
	node[white] {{\fontfamily{qag}\selectfont \tiny ID}};
	\draw[white, fill=white] (-0.0625,0.095)
	circle [radius=0.007];
	\end{tikzpicture}
	\hspace{-2mm}
}
\foreach \x in {A, ..., Z}{\expandafter\xdef\csname orcid\x\endcsname{\noexpand\href{https://orcid.org/\csname orcidauthor\x\endcsname}
			{\noexpand\orcidicon}}
}

\title{A Two-Stage Approach for Segmenting Spatial Point Patterns Applied to Multiplex Imaging}

\usepackage{xwatermark}

\newwatermark[firstpage,color=gray!90,angle=0,scale=0.28, xpos=0in,ypos=-5in]{*correspondence: \texttt{sheng123@umn.edu}}

\usepackage{authblk}
\renewcommand*{\Authfont}{\bfseries}

\begin{document}
\author[1]{Alvin Sheng} 
\author[2]{Brian J Reich} 
\author[2]{Ana-Maria Staicu}
\author[3,4]{Santhoshi N. Krishnan}
\author[3,4,5,6,7]{Arvind Rao}
\author[8]{Timothy L Frankel}

\affil[1]{Division of Biostatistics and Health Data Science, University of Minnesota, Minneapolis, MN, USA}
\affil[2]{Department of Statistics, North Carolina State University, Raleigh, NC, USA}
\affil[3]{Department of Computational Medicine and Bioinformatics, University of Michigan, Ann Arbor, MI, USA}
\affil[4]{Department of Electrical and Computer Engineering, Rice University, Houston, TX, USA}
\affil[5]{Department of Biostatistics, University of Michigan, Ann Arbor, MI, USA}
\affil[6]{Department of Radiation Oncology, University of Michigan, Ann Arbor, MI, USA}
\affil[7]{Department of Biomedical Engineering, University of Michigan, Ann Arbor, MI, USA}
\affil[8]{Department of Surgery, University of Michigan, Ann Arbor, MI, USA}


\twocolumn[ 
  \begin{@twocolumnfalse} 

\maketitle

\begin{abstract}
Recent advances in multiplex imaging have enabled researchers to locate different types of cells within a tissue sample.
This is especially relevant for tumor immunology, as clinical regimes corresponding to different stages of disease or responses to treatment may manifest as different spatial arrangements of tumor and immune cells. 
Spatial point pattern modeling can be used to partition multiplex tissue images according to these regimes. To this end, we propose a two-stage approach: first, local intensities and pair correlation functions are estimated from the spatial point pattern of cells within each image, and the pair correlation functions are reduced in dimension via spectral decomposition of the covariance function. Second, the estimates are clustered in a Bayesian hierarchical model with spatially-dependent cluster labels. The clusters correspond to regimes of interest that are present across subjects; the cluster labels segment the spatial point patterns according to those regimes. Through Markov Chain Monte Carlo sampling, we jointly estimate and quantify uncertainty in the cluster assignment and spatial characteristics of each cluster. Simulations demonstrate the performance of the method, and it is applied to a set of multiplex immunofluorescence images of diseased pancreatic tissue.
\end{abstract}
\vspace{0.35cm}

  \end{@twocolumnfalse} 
] 



\section{Introduction}
\label{sec:intro}

Tissue images are conventionally analyzed with a staining technique like immunohistochemistry, which labels the instances of a single marker in a tissue sample where the marker corresponds to a specific cell phenotype \citep{krishnan_gawrdenmap_2022}. However, recent breakthroughs in multiplex imaging have allowed researchers to simultaneously visualize and quantify multiple markers in a single tissue sample \citep{tan2020overview}. Multiplex imaging technology has been used within the field of tumor immunology to deepen our understanding of the cellular interactions within the tumor microenvironment (TME), or the complex network of cells and other structures surrounding a tumor. Studying the TME is pivotal for understanding how cancer cells can grow, metastasize, and resist drug therapy \citep{anderson_tumor_2020}. Thus, spatial analysis of multiplex imaging data may offer insight into how cellular interactions within the TME affect cancer prognoses and responses to treatment. In the tumor immunology literature, there have been many successful attempts to garner insights from tissue images by analyzing them as spatial point patterns (SPP), where cells are represented as points with different types; for example, see \citet{Lazarus:2018wu}, \citet{wang_rcnn}, and \citet{steinhart_spatial_2021}. 


Specifically, we are motivated by the task of deriving insights from a set of SPPs retrieved from multiplex immunofluorescence (mIF) images of diseased pancreatic tissue. These images were obtained from patients at the University of Michigan Pancreatic Cancer Clinic that were diagnosed into one of six pancreatic disease groups. This motivating dataset was recently used to compare the spatial and phenotypic characteristics of the six disease groups \citep{enzler_comparison_2024}. Some co-authors of this paper have also used this dataset to motivate several supervised machine learning algorithms for SPPs where the pancreatic disease groups are the labels. These include classification algorithms based on neural nets \citep{li2021srnet,baranwal2021cgat}, geographically weighted regression \citep{krishnan_gawrdenmap_2022}, and dictionary learning \citep{krishnan_towards_2023}, as well as an algorithm to discover contrasting co-location patterns between two disease groups \citep{li_cscd_2022}.

In contrast to the previous methods motivated by the mIF images of pancreatic tissue, we propose an unsupervised clustering algorithm to segment the mIF images on the basis of SPP analysis.
Note that in the tissue imaging literature, ``segmentation" often refers to the delineation of the boundaries of cells or nuclei within the image. 
However, we use the word ``segmentation" to refer to partitioning the SPP retrieved from the image. 
Segmenting a multiplex image into subregions with distinct spatial characteristics may reveal different regimes of clinical significance, e.g., different stages of disease.

Several methods for partitioning SPPs have been developed, which generally take one of two approaches. The first is the model-based approach, which fits a mixture model to the point locations and then uses the model to estimate the regime membership of each point. 
Most model-based methods are designed for only two regimes of spatial characteristics: i) ``clutter", or points representing random noise, and ii) ``feature", or points with a spatial distribution distinct from that of clutter. For instance, the feature points may be more equally-spaced or have a higher intensity, i.e., expected number of points per unit area. Some methods that fall in this category are \citet{dasgupta_raftery}, \citet{byers_raftery}, and \citet{cressie_hierarchical_2000}. These methods typically rely on strong assumptions regarding the SPP.

The second approach for partitioning SPPs is the two-stage approach, which first computes localized summary statistics of the SPPs and then clusters the statistics. 
Many two-stage methods rely on the local indicator of spatial association (LISA) proposed by \citet{anselin1995local}, which is a localized version of a global estimator of spatial characteristics. 
For instance, \citet{cressie_collins} compute the product density LISA function for each point and cluster the points based on the distance between functions. \citet{gonzalez_classification_2021} use the same approach, but with pair correlation LISA functions. 
Additionally, the R package \texttt{lisaClust} computes the L-function LISA for each point, evaluates it at a set of values, and clusters the resulting vectors via $k$-means \citep{lisaClust}. Two-stage methods like those mentioned above offer a more flexible, nonparametric approach to clustering as compared to the model-based approach.

The model-based and two-stage approaches mentioned above partition the SPP by classifying the points into regimes. 
In contrast, \citet{allard_nonparametric_1997} propose a model-based approach which partitions the SPP by estimating the subregions as unions of Voronoï polygons. 
As \citet{allard_nonparametric_1997} point out, segmenting the domain according to certain regimes instead of only classifying each point is valuable in situations where the area of influence around a point has some significance. This is especially the case in tumor immunology, where every cell has a potential influence on the surrounding TME. Because of the high intensity of cells in most multiplex images, we propose to subdivide each image with a uniform grid instead of Voronoï polygons.

We build upon previous work on partitioning SPPs by proposing a novel two-stage approach for segmenting SPPs arising from multiplex images. In the first stage, we set each SPP upon a grid and compute local estimates of first- and second-order spatial characteristics for each grid region. 
The local estimates of the second-order spatial characteristics are functional; we use the spectral decomposition of their covariance function to extract their low-dimensional features.
In the second stage, we cluster these low-dimensional features with a Bayesian hierarchical clustering model that uses a Potts model prior, henceforth called the ``Potts clustering model" (PCM), to find subregions corresponding to biologically interpretable regimes. Unlike model-based approaches, our two-stage approach has less restrictive assumptions on the SPP process and intensity, only assuming local homogeneity within a grid region. Unlike the other two-stage approaches previously discussed, our method focuses on segmenting the domain of each SPP rather than classifying each point, which is more appropriate for multiplex images with high intensities of points. 

The PCM in the second stage of our method has a couple of desirable features. First, it is able to discover biologically interpretable and visually informative regimes from tissue images. 
Additionally, unlike the previously discussed methods for partitioning SPPs, the PCM in the second stage can incorporate subject-specific categorical data and pool information across multiple subjects to more accurately estimate each regime's spatial characteristics.

The paper is organized as follows. 
In Section \ref{sec:model_framework}, we discuss the spatial statistics that our two-stage approach relies on, as well as the theoretical assumptions of each stage of our approach. In Section \ref{sec:estimation}, we discuss the estimation of the spatial statistics and other computational details. In Section \ref{sec:sim_sensitan}, we validate our approach via a simulation study. In Section \ref{sec:real_data}, we apply our approach to a dataset of mIF images of diseased pancreatic tissue. Finally, in Section \ref{sec:discuss}, we end with some conclusions.

\section{Model Framework}
\label{sec:model_framework}

We assume that the data observed for subject $n = 1, \ldots, N$ are

\begin{eqnarray} \label{eq:spp_sn}
    S_n = \{ (\bs_{ni}, \tau_{ni})_{i = 1, \ldots, m_n} \},
\end{eqnarray}

\noindent where $\bs_{ni} = (s_{ni1}, s_{ni2}) \in W_n \subset \mathbb{R}^2$ is the spatial location and $\tau_{ni} \in \{ 1, \ldots, H \}$ is the known cell type of cell $i$ in subject $n$. Here $W_n$ denotes the sampling window for subject $n$ and $|W_n|$ its area. The number of cells, their locations, and types are modeled as an SPP.

We assume that the local heterogeneity of each SPP is explained by a partition of homogeneous subregions that correspond to biologically interpretable regimes. We also assume that the subregions are well described by combinations of grid rectangles, or grid regions. To uncover these subregions, we propose a two-stage approach. In the first stage, we set each SPP upon a grid and extract estimates of the local first- and second-order characteristics of each grid region. Web Appendix A defines the first- and second-order characteristics used, namely, the intensity and pair correlation function (PCF). In the second stage, we cluster these local characteristics with the PCM to find the subregions corresponding to each regime. In this paper, ``cluster" and ``regime" are synonymous.

\begin{figure*}
\centering
  \includegraphics[width=0.9\textwidth]{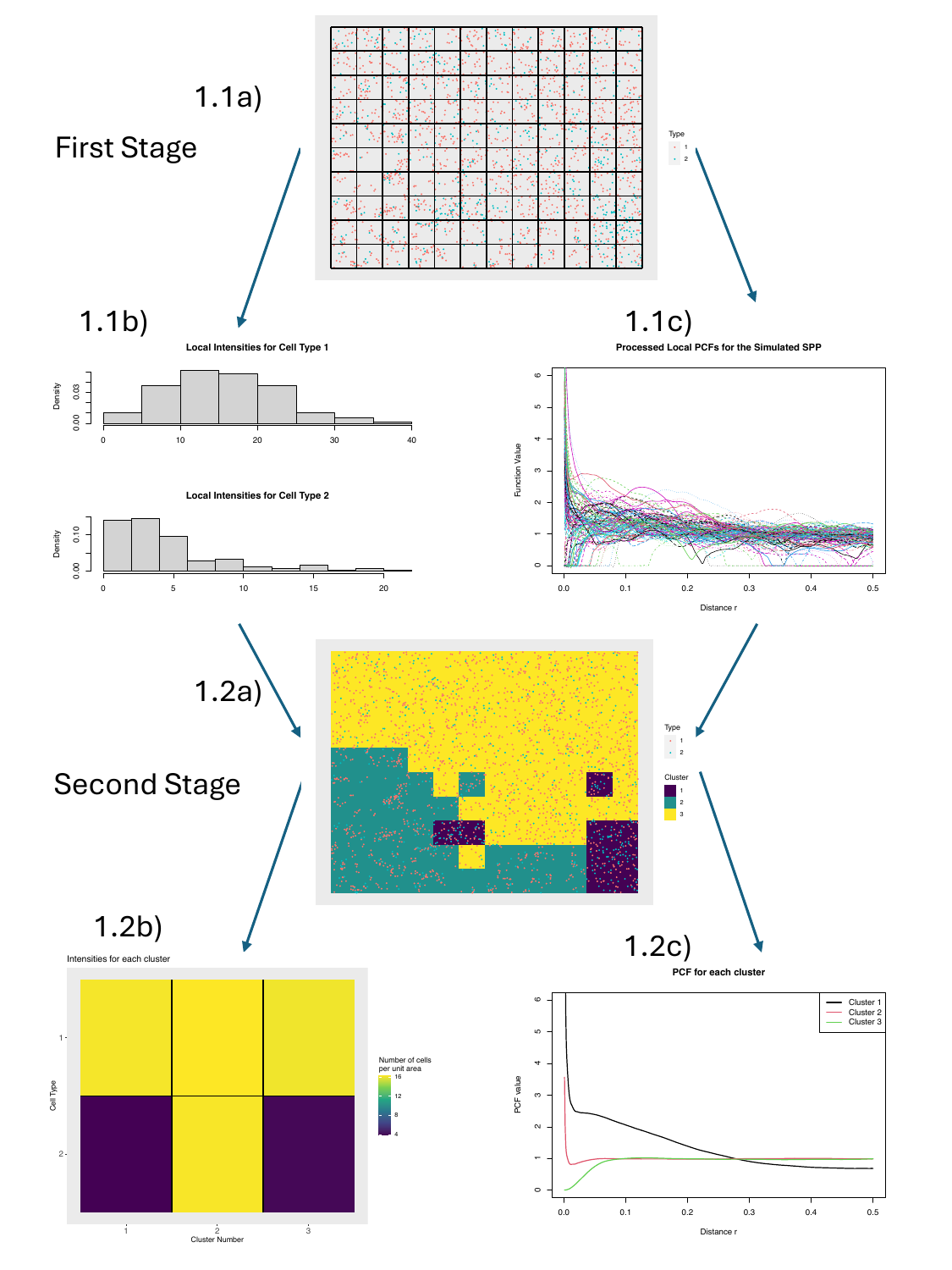}
\caption{A schematic of our two-stage approach, using an example SPP. The top half of the schematic represents the first stage, where the SPP is set upon a grid and the local intensities and PCFs are estimated for each grid region. The bottom half of the schematic represents the second stage, where the grid regions are clustered on the basis of the local estimates and the regimes' spatial characteristics are estimated and summarized.}
\label{fig:schematic_two_stage}
\end{figure*}

Each stage of the approach is summarized by a schematic in Figure \ref{fig:schematic_two_stage}. Sections \ref{local} and \ref{bhcm} describe the modeling assumptions of the first and second stage, respectively; Section \ref{sec:estimation} discusses the estimation procedure for the two-stage approach. 

\subsection{First stage: local features of the SPPs}
\label{local}

We subdivide each SPP $S_n$ from Equation \ref{eq:spp_sn} according to a given grid (fixed a priori) based on the centers $\{ \bx_1, \ldots, \bx_{L} \}$ that span $W_n$:

\begin{eqnarray}
W_{nl} = \{ \bs \in W_n: \argmin_k ||\bs - \bx_{k}|| = l \}, \ l = 1, \ldots, L.
\end{eqnarray}

\noindent Then, each grid region $W_{nl}$ contains a subset of $S_n$, or a subcollection of cell locations $\bs$ and types $\tau$ from $S_n$: 

\begin{eqnarray}
    S_{nl} = \{ (\bs, \tau) \in S_n : \bs \in W_{nl} \}, \ l = 1, \ldots, L.
\end{eqnarray}

We assume the grid regions are sufficiently small so that the process governing the first-order characteristic can be modelled as locally homogeneous within the grid regions. That is, the local intensity for type $h$ and local marginal intensity, respectively, are defined as $\lambda_{nlh} = \lambda_{nh}(\bx_l)$ and $\lambda_{nl} = \lambda_{n}(\bx_l)$. See Web Appendix A for definitions of the intensity functions $\lambda_{nh}(\cdot)$ and $\lambda_{n}(\cdot)$.

Similarly, assume that the process governing the second-order characteristic is locally homogeneous and isotropic. Then, the local product density function depends only on the distance between points $\bs, \bs' \in W_{nl}$, rather than the locations of the two points. By an abuse of notation we define the local product density function as $\lambda^{(2)}_{nl}(\bs, \bs') = \lambda^{(2)}_{nl}(r)$, where $\bs, \bs' \in W_{nl}$ and $r = ||\bs - \bs'||$. See Web Appendix A for the definition of the product density function. The corresponding local PCF for grid region $l$ is then defined as 

\begin{eqnarray}
    g_{nl}(r) = \frac{\lambda_{nl}^{(2)}(r)}{\lambda_{nl}^2}.
\end{eqnarray}

We truncate the domain of the local PCF $g_{nl}(r)$ at half the shortest dimension of a grid region, which we denote as $R$. We truncate the local PCF at this value to capture only the local information, which is the most relevant biologically. Additionally, the PCF has high variability for larger distances.
To reduce the skewness of the function values, we also take the square root of the truncated local PCF. Denote the truncated and transformed function as $X_{nl}(r) = \sqrt{g_{nl}(r)}$ for $r \in [a, R]$, where $a$ is some positive value less than $R$.

Denote by $M$ the total number of regimes across subjects and let $C_{nl} \in \{ 1, \ldots, M \}$ be the regime assignment for subject $n$ and grid region $l$. We assume that the number of regimes $M$ is fixed or pre-specified, but the regime labels $C_{nl}$ are unknown and distributed according to the Potts model discussed in Section \ref{bhcm}. All grid regions assigned to regime $\eta$ across subjects have the same local intensities $\{ \lambda_{h \eta} \}_{h = 1}^H$ and local PCF $g_{\eta}(\cdot)$ for that regime. Because the local PCF $g_{nl}(\cdot) = g_{\eta}(\cdot)$ if $\eta = C_{nl}$, we have that $X_{nl}(r) = \sum_{\eta = 1}^M \bI(C_{nl} = \eta) \sqrt{g_{\eta}(r)}, \ r \in [a, R]$. $X_{nl}(\cdot)$ is a random function that is completely determined by the random assignment of the cluster label $C_{nl}$.


We extract important features of these curves by using the eigenfunctions of the spectral decomposition of the covariance function, in a manner similar to functional principal component analysis. Let $\mu(r)$ and $\Sigma(r, r')$ be the marginal mean and covariance functions of $\{ X_{nl}(\cdot) \}^{1 \leq n \leq N}_{1 \leq l \leq L}$ over the cluster labels $C_{nl}$. 
The spectral decomposition of $\Sigma(r, r')$ yields the pairs of eigenvalues and orthogonal eigenfunctions $\{ \rho_k, \phi_k(r) \}_{k \geq 1}$ for $\rho_1 \geq \rho_2 \geq \ldots \geq 0$. We use the leading eigenfunctions $\phi_k(\cdot)$'s to extract the key features of $X_{nl}(\cdot)$ that we call ``scores." Specifically, the $k^{th}$ score is defined by

\begin{eqnarray} \label{pc_proj}
    \xi_{nlk} = \int_0^R \{X_{nl}(r) - \mu(r)\} \phi_k(r) dr.
\end{eqnarray}

\noindent The scores $\{ \xi_{nlk} \}_{k \geq 1}$ are analogous to the ``functional principal component scores" associated with the main directions (i.e., eigenfunctions from the spectral decomposition) present in functional data literature \citep{kokoszka_introduction_2017}. However, what we have described is not the typical functional principal component analysis, because the curves $\{ X_{nl}(\cdot) \}_{1 \leq l \leq L}^{1 \leq n \leq N}$ are neither independent nor identically distributed. 

Let $K$ be the number of main directions $\phi_k(\cdot)$ that is sufficient to explain a preset percentage of variance. Then $X_{nl}(\cdot)$ can be approximated by the finite truncation

\begin{align}
    X_{nl}^K(r) = \mu(r) + \sum_{k = 1}^K \phi_k(r) \xi_{nlk}.
\end{align}

\noindent The scores $\{ \xi_{nl1}, \ldots, \xi_{nlK} \}$ can be interpreted as the set of important features extracted from $X_{nl}(\cdot)$; they are parameters that describe the second-order characteristic of each grid region. We append the intensities $(\lambda_{nl1}, \ldots, \lambda_{nlH})$ to the vector of scores by defining $\xi_{nlK+h} \coloneqq \lambda_{nlh}$, giving $Q = K + H$ values $\bxi_{nl} = (\xi_{nl1}, \ldots, \xi_{nlK}, \xi_{nlK+1}, \ldots, \xi_{nlQ})^\intercal \in \mathbb{R}^{Q}$ for subject $n$ in grid region $l$. In Section \ref{sec:estimation}, we describe how the scores and intensities in $\bxi_{nl}$ are estimated from the data. Then, $\bxi_{nl}$ is centered and scaled to have mean 0 and standard deviation 1 across grid regions and subjects. The array of parameters $\bXi = \{ \bxi_{nl} \}_{1 \leq n \leq N}^{1 \leq l \leq L}$ is then clustered in the second stage to find regimes of spatial characteristics across subjects.

\subsection{Second stage: Potts clustering model}
\label{bhcm}

The second stage is the PCM which uses the extracted features $\bxi_{nl}$ for all subjects $n$ and grid regions $l$. The proposed model is a mixture model with spatially dependent cluster labels that partition the images into spatially coherent subregions. We denote the centered and scaled scores and intensities corresponding to regime $\eta$ across subjects as $\mu_{q \eta}, q = 1, \ldots Q$. Then, 

\begin{align} \label{theo_first_layer}
    \xi_{nlq} = \sum_{\eta = 1}^M \bI(C_{nl} = \eta) \mu_{q \eta},
\end{align}

\noindent for $n = 1, \ldots, N$, $l = 1, \ldots, L$, and $q = 1, \ldots, Q$. The parameters $\{ \mu_{q \eta} \}_{q = 1}^{Q}$ govern the first- and second-order characteristics of subregions in regime $\eta$.

For a given subject, we assume that each grid region is likely to share the same regime as neighboring grid regions. We use the spatial Potts model for the cluster labels \citep*{Gelfand:2014aa} to formalize this and represent the full conditional probability for $C_{nl}$ as

\begin{align}
        P(C_{nl} = \eta | C_{nl'}, &l' \neq l) \propto \nonumber \\ &\exp \left \{ \alpha_\eta + \psi \sum_{l' \in \mathcal{N}_{nl}} \bI(C_{nl'} = \eta) \right \},
\end{align} 

\noindent where $\mathcal{N}_{nl}$ is the set of indices $l'$ such that grid region $W_{nl'}$ is adjacent (horizontally or vertically) to grid region $W_{nl}$, $\alpha_\eta$ is the offset determining the proportion of grid regions in subject $n$ assigned to cluster $\eta$, and $\psi \geq 0$ is the smoothness parameter determining the strength of spatial dependence between adjacent grid regions. These conditional distributions imply the joint probability mass function

\begin{equation}
    p(C_{n1}, \ldots, C_{nL}|\btheta) =
    \frac{1}{d(\btheta)} \exp{\left\{z(C_{n1}, \ldots, C_{nL}, \btheta) \right\}} \label{spatial_potts}
\end{equation}

\begin{align}
    z(C_{n1}, \ldots, C_{nL}, \btheta) = &\sum_{\eta = 1}^M \alpha_\eta \left\{\sum_{l = 1}^L \bI(C_{nl} = \eta) \right\} + \nonumber \\ &\psi \sum_{l \sim l'} \bI(C_{nl} = C_{nl'}),
    \label{z_eqn1}
\end{align}

\begin{align}
    d(\btheta) = \sum_{C_{n1} = 1}^M \cdots \sum_{C_{nL} = 1}^M
    \exp{\left\{ z(C_{n1}, \ldots, C_{nL}, \btheta) \right\}} \label{d_theta}
\end{align}

\noindent for $n = 1, \ldots, N,$ where $\btheta = (\psi, \alpha_1, \ldots, \alpha_M)$ and $l \sim l'$ indicates that grid regions $W_{nl}$ and $W_{nl'}$ are adjacent. To make the spatial Potts parameters identifiable, we set $\alpha_1$ to 0. 

The normalizing constant $d(\btheta)$ is intractable since it involves a large number $M^L$ of terms, but is required for implementing the PCM. To evaluate the normalizing constant, we use an approach suggested by \citet{Reich:2014aa}, which involves estimating expected ``sufficient quantities" via simulation and interpolating among them via a machine learning algorithm (more details are given in Web Appendix B.1.1). We define sufficient quantities to be values that capture all the information about the spatial Potts parameters contained in the cluster labels. The sufficient quantity for $\alpha_\eta$ is the number of grid regions in the subject assigned to cluster $\eta$, $\sum_{l = 1}^L \bI(C_{nl} = \eta)$, and the sufficient quantity for $\psi$ is the number of matching adjacent pairs in the subject, $\sum_{l \sim l'} \bI(C_{nl} = C_{nl'})$

We specify noninformative priors for the other parameters. Because the parameter vectors $\{ \bxi_{nl} \}_{1 \leq n \leq N}^{1 \leq l \leq L}$ have been centered and scaled, we also center and scale the priors for $\mu_{q \eta}$ such that $\mu_{q \eta} \overset{iid}{\sim} N(0, 1)$ for $q = 1, \ldots, Q$ and $\eta = 1, \ldots, M$. Thus, the distributions of both the grid region parameter $\xi_{nlq}$ and regime parameter $\mu_{q \eta}$ are centered at 0 with standard deviation 1. To span the range of plausible $\alpha_\eta$ and $\psi$ values, the priors for the spatial Potts model parameters are specified as $\alpha_\eta \sim Uniform(-5, 5)$ for $\eta = 2, \ldots, M$ and $\psi \sim Uniform(0, 2.5)$. The minimum and maximum values in the above ranges for $\alpha_\eta$ and $\psi$ are those that approximately achieve the minimum and maximum expected values of their respective sufficient quantities.

In summary, $C_{nl}, \mu_{q \eta}$, and $\btheta$ are the unknown hyperparameters of interest. By examining the posterior distribution of these hyperparameters, we can infer the shapes, sizes, and spatial characteristics of various regimes of potential clinical significance. The computational details for this inference and summarizing the posterior distribution are discussed in the next section and Web Appendix B, respectively.

\section{Computational details}
\label{sec:estimation}

\begin{figure}
\centering
  \includegraphics[width=0.5\textwidth]{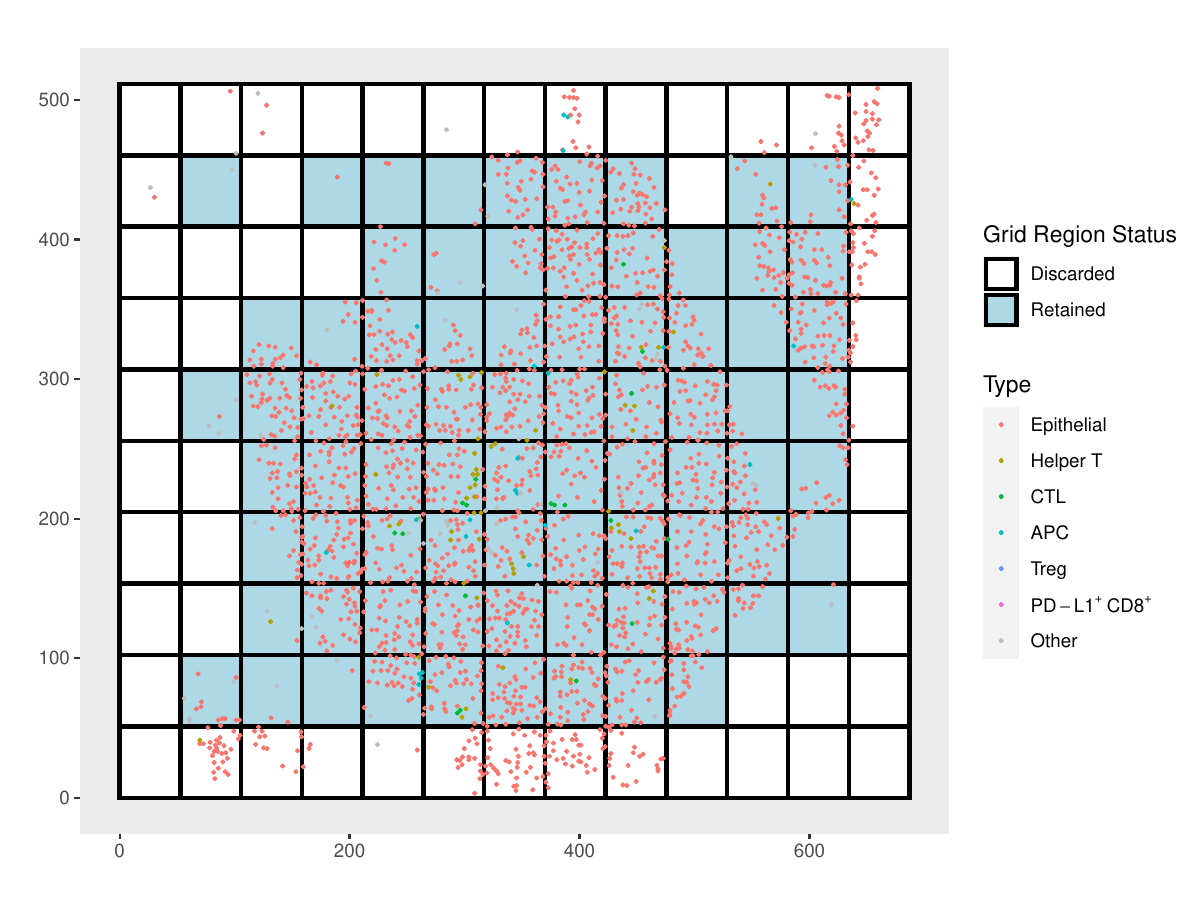}
\caption{A representative SPP from the motivating dataset (see Web Appendix E for a detailed description) with a $10 \times 13$ grid overlaid. The grid regions that have at least one cell and are fully within the window of the SPP are retained (marked in blue), whereas other grid regions are discarded (marked in white).}
\label{fig:gridded_spp}
\end{figure}

First, we divide each SPP with the same rectangular grid such that each grid region has an average of 20 cells across subjects. This ensures that the grid regions have sufficient data to make differences between regimes distinguishable, while being small enough to justify the local homogeneity assumption. Grid regions should have the same size and have at least one cell; grid regions $W_{nl}$ that cross the boundary of $W_n$ or contain no cells are discarded, and statistics associated with them are set as missing. Furthermore, grid regions that contain less than two cells have their PCF statistics set as missing. Figure \ref{fig:gridded_spp} shows an example of a gridded SPP from the motivating dataset, with some grid regions discarded. In Web Appendix D.1, we examine the robustness of the method's performance under different grid resolutions through a sensitivity analysis.



The local intensity for each cell type in a grid region is estimated to be the number of points per unit area

\begin{align} \label{eq:local_intens}
    \hat{\lambda}_{nlh} &= \frac{|\{ (\bs, \tau) \in S_{nl}: \tau = h \}|}{|W_{nl}|}, \ h = 1, \ldots, H.
\end{align}

\noindent This is an unbiased estimate of the true intensity $\lambda_{nlh}$, assuming that the point process has homogeneous intensity in the grid region \cite*[Chapter 6]{Baddeley:2016aa}. See Figure 1.1b for an example of the estimated local intensities for a representative subject. Likewise, the local marginal intensity is $\hat{\lambda}_{nl} = |\{ (\bs, \tau) \in S_{nl} \}| / |W_{nl}|$, which is an unbiased intensity of $\lambda_{nl}$.

To estimate the local PCF for the grid region, we use the fixed-bandwidth kernel estimator \citep[Chapter 15]{stoyan1994fractals}:

\begin{align}
    \hat{g}_{nl}(r) = &\frac{|W_{nl}|}{2 \pi r m_{nl} (m_{nl} - 1)} \times \nonumber \\ &\sum_{i = 1}^{m_{nl}} \sum_{j = 1, j \neq i}^{m_{nl}} \epsilon_w(r - ||v_{nl,ij}||) t(v_{nl,ij}, W_{nl}) \label{local_pcf} \\
    t(v, W) = &\frac{|W|}{|W \cap (W - v)|}, \label{transcorrection}
\end{align}


\noindent for $n = 1, \ldots, N$ and $l = 1, \ldots, L$, where $m_{nl}$ is the number of cells in $W_{nl}$, $i, j$ are indices of cells in $W_{nl}$, $v_{nl,ij} = \pmb s_{nlj} - \pmb s_{nli}$, $\epsilon_w(\cdot)$ is the Epanechnikov smoothing kernel with the half-width $w > 0$, and $t(v, W)$ is the translation edge correction. Because the local PCF estimator only uses points within the grid region $W_{nl}$, edge effects may bias the estimator if not addressed. The edge correction accounts for edge effects by applying the Horvitz-Thompson principle \citep{ohser1983estimators,ohser1981second}. The edge correction is valid for distances $r \leq 2R$, where $2 R$ is the length of the shortest dimension of the grid region \citep[Chapter 7]{Baddeley:2016aa}. We select the half-width $w$ for the Epanechnikov kernel according to Stoyan's rule of thumb, which sets $w = \frac{0.15}{\sqrt{m_{nl} / |W_{nl}|}}$ \citep[Chapter 15]{stoyan1994fractals}.

The local PCF estimator $\hat{g}_{nl}(r)$ is processed by taking the square root and fitting a cubic smoothing spline, leading to the estimate $\hat{X}_{nl}(r)$. Figure 1.1c shows the set of processed curves $\{ \hat{X}_{nl}(r) \}_{l = 1}^{130}$ calculated for a representative image.  
We can extract important features of the processed curves by using the eigenfunctions of the spectral decomposition of the sample covariance function, in the manner of Section \ref{local}. Here, we estimate $\mu(r)$ by $\bar{X}(r)$, the average of $\hat{X}_{nl}(r)$ over $n, l$, and $\Sigma(r, r')$ by $\hat{\Sigma}(r, r'),$ the sample covariance function. Spectral decomposition of $\hat{\Sigma}(r, r')$ yields the pairs of eigenvalues and orthogonal eigenfunctions $\{ \hat{\rho}_k, \hat{\phi}_k(r) \}_{k \geq 1}$ for $\hat{\rho}_1 \geq \hat{\rho}_2 \geq \ldots \geq 0$. We use the leading eigenfunctions $\hat{\phi}_k(\cdot)$'s to compute the scores of $\hat{X}_{nl}(\cdot)$:

\begin{eqnarray} \label{pc_proj_est}
    \tilde{\xi}_{nlk} = \int_0^R \{\hat{X}_{nl}(r) - \bar{X}(r)\} \hat{\phi}_k(r) dr.
\end{eqnarray}

In practice, we discretize the processed curves by denoting $\hat{X}_{nl} \in \mathbb{R}^{R_d}$ as the vector of $\hat{X}_{nl}(r)$ values evaluated at $r  = \frac{R}{R_d}, \ldots, R$, where $R_d$ is given. We then apply the algorithm of principal component analysis on the vectors of discretized function values $\{ \hat{X}_{nl} \}_{1\leq l \leq L}^{1 \leq n \leq N}$ (with the mean vector $\bar{X}$), yielding the pairs of eigenvalues and orthogonal eigenvectors $\{\hat{\rho}^*_k, \hat{\phi}_k\}_{k = 1}^{R_d}$ for $\hat{\rho}^*_1 \geq \hat{\rho}^*_2 \geq \ldots \geq \hat{\rho}^*_{R_d}$. Then, we instead focus on the scores for $\hat{X}_{nl}$: 

\begin{align} \label{pc_proj_riemann}
    \hat{\xi}_{nlk} = (\hat{X}_{nl} - \bar{X})^\intercal \hat{\phi}_k.
\end{align}

\noindent This is analogous to performing functional principal component analysis by discretizing the functional data and then performing principal component analysis \citep[Chapter 8]{Ramsay:2005vf}.

Let $K$ be the number of eigenvectors $\hat{\phi}_k$ that explain 80\% of the variance in $\{ \hat{X}_{nl} \}_{1\leq l \leq L}^{1 \leq n \leq N}$, according to the eigenvalues $\hat{\rho}^*_k$. 
Then $\hat{X}_{nl}$ can be approximated by the truncation

\begin{align}
    \hat{X}^K_{nl} = \bar{X} + \sum_{k = 1}^{K} \hat{\phi}_k \hat{\xi}_{nlk}.
    \label{eq:empirical_finite_truncation}
\end{align}

\noindent We append the intensities $(\hat{\lambda}_{nl1}, \ldots, \hat{\lambda}_{nlH})$ to the vector of estimated scores by defining $\hat{\xi}_{nlK+h} \coloneqq \hat{\lambda}_{nlh}$, giving $Q = K + H$ values $\hat{\bxi}_{nl} = (\hat{\xi}_{nl1}, \ldots, \hat{\xi}_{nlK}, \hat{\xi}_{nlK+1}, \ldots, \hat{\xi}_{nlQ}) \in \mathbb{R}^{Q}$ for subject $n$ in grid region $l$. For numerical stability, we center and scale the components of $\hat{\bxi}_{nl}$ to have mean 0 and standard deviation 1. The array of estimates $\hat{\bXi} = \{ \hat{\bxi}_{nl} \}_{1 \leq n \leq N}^{1 \leq l \leq L}$ is then clustered in the second stage to find regimes of spatial characteristics across subjects.


To account for measurement error, we adjust the PCM in the second stage of our approach to include the error variance $\nu^2_q, q = 1, \ldots, Q,$ in the first layer as follows:

\begin{align} \label{first_layer}
    \hat{\xi}_{nlq}|\mu_{q1}, \ldots, \mu_{qM}, C_{nl}, \nu^2_{q} \indep \nonumber \\ N\left(\sum_{\eta = 1}^M \bI(C_{nl} = \eta) \mu_{q \eta}, \  \nu_{q}^2 \right),
\end{align}

\noindent for $n = 1, \ldots, N$, $l = 1, \ldots, L$, and $q = 1, \ldots, Q$. Compare this with (\ref{theo_first_layer}), which would be the case without measurement error. The noninformative priors for the additional parameters $\nu_q^2$ are

\begin{align} \label{nu_prior}
    \nu_q^2 \overset{iid}{\sim} InvGamma(1, 0.01),
\end{align}

\noindent for $q = 1, \ldots Q$. The other layers of the PCM are the same as in Section \ref{bhcm}. 

To conduct inference with the PCM, we implement Markov Chain Monte Carlo (MCMC) sampling in R \citep{Rsoftware} to approximate the posterior distribution of the cluster labels and regime-specific parameters. We discuss the MCMC sampling algorithm in Web Appendix B.1 and posterior inference in Web Appendix B.2.

\section{Simulation study}
\label{sec:sim_sensitan}

In this section, we evaluate the performance of the proposed method with a simulation study and sensitivity analyses. We first introduce competing algorithms, then describe the simulation scenarios, and finally present the results.

\subsection{Competing algorithms and metrics}
\label{sec:competing_alg}

Like the PCM, the competing algorithms will have as the input the processed local intensity and PCF estimates in $\hat{\bXi}$, computed by the first stage of our approach on generated SPPs (Section \ref{sec:estimation}). In the first stage, we compute local estimates according to a $10 \times 12$ grid, where each grid region is a unit square (however, the sensitivity analysis in Web Appendix D.1 will examine the effect of using different grid resolutions). We discretize the local PCF estimates at $R_d = 512$ equispaced r-values from 0.001 to 0.5 units to get the vectors $\hat{X}_{nl}$.

The first competing algorithm is the PCM where the spatial smoothness parameter is set to zero (``non-spatial PCM"). The PCM and non-spatial PCM will be run for 30,000 iterations, discarding the first 10,000 as burn-in. The other competing algorithms are four variations of the $k$-means algorithm of \citet{hartigan1979algorithm}. The first point of difference between variations is whether the discretized function values $\hat{X}_{nl}$ are reduced via principal component analysis or left as is before being appended to the intensity estimates. In the former case, the inputs $\hat{\bXi}$ would consist of $(K + H)$-dimensional vectors as described in Section \ref{sec:estimation}, where $K$ is the number of eigenvectors retained and $H$ is the number of cell types (denote this variation with the prefix ``FPCA"); in the latter case, $\hat{\bXi}$ would consist of $(R_d + H)$-dimensional vectors (denote this variation with the prefix ``Curve"). In both cases, the vectors in $\hat{\bXi}$ are centered and scaled to have mean 0 and standard deviation 1. The second point of difference between variations is whether the $k$-means algorithm is applied to all subjects globally or to each subject separately. In the former case, the $k$-means algorithm is applied to $\hat{\bXi}$ directly (denote this variation with the suffix ``G"). In the latter case, the $k$-means algorithm is applied to $\{ \hat{\bxi}_{nl} \}_{l = 1}^L$, separately for each $n = 1, \ldots, N$ (denote this variation with the suffix ``S"). The average performance of the $N$ clustering results is considered. The four $k$-means variations are denoted as FPCA-G, FPCA-S, Curve-G, and Curve-S. The correct number of clusters is specified a priori for all methods in the simulation study, except for the sensitivity analysis regarding the selection of different numbers of clusters (Web Appendix D.2).

The performance of each method is evaluated by comparing the method's partitioning of the grid regions with the true partitioning. The comparison is done using the Adjusted Rand Index (ARI) \citep{Hubert:1985uo}, which measures the similarity between two partitions by counting pairs of elements that are clustered together or separately in the two partitions. The two partitions may have different numbers of clusters. If the two partitions are almost identical, the ARI is close to one; if one of the partitions was randomly generated, the ARI tends toward zero. The ARI can be negative if the proposed partition is worse than a random partition. The ARIs for each scenario in the simulation study and sensitivity analyses will be averaged over 100 Monte Carlo replications.

\subsection{Data generation and simulation study results}
\label{sec:main_sim_study}

The simulation study examines the performance of the PCM and competing methods in 16 scenarios determined by four factors, under which the SPPs are generated on a $10 \times 12$ grid. The factors listed, with the factor levels in parantheses, are i) the number of clusters $M$ (three, five), ii) smoothness parameter $\psi$ (0, 1.29), iii) number of subjects $N$ (50, 100), and iv) total intensity $\lambda$ of cells (low, high). In regards to the fourth factor, we generate two types of cells, whose total intensity $\lambda$ ranges from 20 to 32 in the low intensity case and 45 to 72 in the high intensity case. We will examine one additional scenario with four clusters in the resolution sensitivity analysis (Web Appendix D.1). Additional details and figures describing the data generation are provided in Web Appendix C.

For the three-cluster scenarios, in the spatial independence case ($\psi = 0$), both the PCM and non-spatial PCM slightly lag behind the competing algorithms FPCA-G and FPCA-S when there is a low intensity, but outperforms the competing algorithms when there is a high intensity (first part of Table \ref{tab:sim_results}). In the spatial dependence case ($\psi = 1.29$), the PCM outperforms the non-spatial PCM and competing algorithms with either the low or high intensity; with the high intensity, the clustering performance of the PCM is nearly perfect (second part of Table \ref{tab:sim_results}).

\begin{table*}[]
\centering
\begin{tabular}{lllllllllll}
\hline
& $M$ & $\psi$ & $N$ & $\lambda$ & PCM & non-spatial & FPCA-G & FPCA-S & Curve-G & Curve-S \\
\hline
\textit{1} & 3 & 0 & 50 & low & 45 (3) & 46 (4) & \textbf{56} (6) & 46 (2) & 20 (1) & 19 (1) \\
& 3 & 0 & 50 & high & 78 (19) & \textbf{79} (18) & 70 (10) & 69 (3) & 24 (1) & 27 (1) \\
& 3 & 0 & 100 & low & 45 (2) & 45 (2) & \textbf{58} (7) & 46 (1) & 20 (1) & 19 (1) \\
& 3 & 0 & 100 & high & \textbf{80} (16) & 79 (17) & 68 (10) & 69 (2) & 24 (1) & 27 (1) \\
\hdashline
\textit{2} & 3 & 1.29 & 50 & low & \textbf{77} (17) & 47 (8) & 55 (8) & 5 (2) & 20 (3) & 3 (1) \\
& 3 & 1.29 & 50 & high & \textbf{96} (8) & 74 (24)  & 69 (14) & 15 (4)  & 25 (3) & 6 (2) \\
& 3 & 1.29 & 100 & low & \textbf{74} (18) & 46 (5) & 55 (7) & 5 (1) & 20 (2) & 3 (1) \\
& 3 & 1.29 & 100 & high & \textbf{97} (8) & 76 (20) & 69 (12) & 16 (2) & 24 (2) & 6 (1) \\
\hdashline
\textit{3} & 5 & 0 & 50 & low & 23 (3) & 23 (3) & \textbf{29} (3) & 29 (1) & 18 (2) & 18 (1) \\
& 5 & 0 & 50 & high & 36 (4) & 36 (3) & \textbf{43} (1) & 42 (1) & 22 (2) & 22 (1) \\
& 5 & 0 & 100 & low & 23 (3) & 23 (3) & \textbf{29} (3) & 29 (1) & 18 (3) & 18 (0) \\
& 5 & 0 & 100 & high & 36 (5) & 37 (3) & \textbf{43} (2) & 42 (1) & 22 (1) & 22 (0) \\
\hdashline
\textit{4} & 5 & 1.29 & 50 & low & \textbf{43} (10) & 23 (3) & 30 (5) & 20 (2) & 18 (2) & 13 (2) \\
& 5 & 1.29 & 50 & high & \textbf{62} (11) & 38 (6) & 44 (5) & 29 (3) & 22 (2) & 17 (2) \\
& 5 & 1.29 & 100 & low & \textbf{43} (11) & 23 (3) & 30 (4) & 20 (2) & 18 (2) & 13 (1) \\
& 5 & 1.29 & 100 & high & \textbf{60} (10) & 36 (6) & 43 (4) & 29 (2) & 22 (2) & 17 (1) \\
\hline
\end{tabular}
\caption{Results of the simulation study. The adjusted Rand index (multiplied by 100) is presented for each of the six competing methods, under sixteen scenarios as determined by four factors. The standard deviation is in parantheses. $M$ is the number of clusters, $\psi$ is the smoothness parameter, $N$ is the number of subjects, and $\lambda$ is the total intensity of two types of cells. Our proposed algorithm is denoted as PCM, whereas the non-spatial PCM is denoted as `non-spatial'. FPCA-G, FPCA-S, Curve-G, and Curve-S are four variations of the $k$-means algorithm. The values of the best performances are in bold type.} \label{tab:sim_results}
\end{table*}

These trends are similar for the five-cluster scenarios; however, since there are five clusters instead of three, all methods tend to achieve lower ARIs. In the spatial independence case, both the PCM and non-spatial PCM have comparable performances with the competing algorithms FPCA-G and FPCA-S, for either intensity (third part of Table \ref{tab:sim_results}). In the spatial dependence case, the PCM outperforms the non-spatial PCM and competing algorithms with either intensity (fourth part of Table \ref{tab:sim_results}).

We also perform sensitivity analyses examining the impact of using different grid resolutions or different numbers of clusters on the methods. Descriptions and results of the sensitivity analyses are presented in Web Appendices D.1 and D.2. We found that using a coarser resolution than the truth does not significantly decrease the performance of the PCM; using more clusters than the truth also does not significantly impact performance.

\section{Analysis of mIF images of pancreatic tissue}
\label{sec:real_data}

In this section, we apply the two-stage approach described in Section \ref{sec:estimation} to the motivating dataset of mIF images of diseased pancreatic tissue. We work with 105 SPPs retrieved from the images, in which there are seven cell types. See Figure \ref{fig:gridded_spp} for a representative SPP, and Web Appendix E for a further description of the motivating dataset. 

For the first stage, we set a $10 \times 13$ grid on the SPPs retrieved from the images. The grid has the same dimensions as the largest window among subjects, which is 687.0 by 511.5 microns. As described in Section \ref{sec:estimation}, for each subject, we discard grid regions from the $10 \times 13$ grid that contain no cells or cross the boundary of the subject's window.
This leads to an average of 22.36 cells per grid region across subjects. We then compute and process the local intensity and PCF estimates for each grid region. We discretize the processed curves at 512 equispaced $r$-values from 0.05 to 25.575 microns. Then, we have the input $\hat{\bXi}$ for the PCM in the second stage.

Because the mIF dataset contains two groups, cancerous and non-cancerous (control) disease, we include additional offsets $\beta_1, \ldots, \beta_M$ in the PCM to allow for different regime, or cluster, assignment probabilities for the two groups. In the adjusted spatial Potts model within the PCM, the cluster labels for subject $n$ have the same probability mass function as shown in Equation \ref{spatial_potts}, but the term in the exponent includes the additional offsets:

\begin{align}
    z(C_{n1}, \ldots, C_{nL}, \btheta) = \sum_{\eta = 1}^M \alpha_\eta \bI_{\text{control}}(n) \left\{\sum_{l = 1}^L \bI(C_{nl} = \eta) \right\} + \nonumber \\ \sum_{\eta = 1}^M \beta_\eta \bI_{\text{cancer}}(n) \left\{\sum_{l = 1}^L \bI(C_{nl} = \eta) \right\} + \psi \sum_{l \sim l'} \bI(C_{nl} = C_{nl'}),
    \label{z_eqn2}
\end{align}

\noindent where $\btheta = (\psi, \alpha_1, \ldots, \alpha_M, \beta_1, \ldots, \beta_M)$, and $\bI_\text{control}(n)$ and $\bI_\text{cancer}(n)$ are indicators for whether subject $n$ is in the control group or the cancer group. The full conditional probability for $C_{nl}$ is then

\begin{align}
        P(C_{nl} = \eta | C_{nl'}, l' \neq l) \propto \exp \biggl \{ &\alpha_\eta \bI_\text{control}(n) + \beta_\eta \bI_\text{cancer}(n) + \nonumber \\
        &\psi \sum_{l' \in \mathcal{N}_{nl}} \bI(C_{nl'} = \eta) \biggr \},
\end{align}

\noindent where $\mathcal{N}_{nl}$ is the set of indices $l'$ such that grid region $W_{nl'}$ is adjacent to grid region $W_{nl}$. To make the spatial Potts parameters identifiable, we set both $\alpha_1$ and $\beta_1$ to 0. We assume that the other parameters $\mu_{q\eta}, \nu^2_q,$ and $\psi$ are the same between the two groups.

Because the PCM takes the number of clusters as fixed, we use the following procedure to select the number of clusters: we fit the PCM for an increasing number of clusters $M = 2, 3, \ldots,$ and stop when the proportion of grid regions in any one cluster is below 1\%. The rationale for this procedure is that beyond a certain number of clusters, the performance of the PCM in terms of the ARI tends not to change very much, according to the sensitivity analysis in Web Appendix D.2. Therefore, it is better to overestimate the number of clusters, even though some clusters may become redundant. For each fitted PCM, we generate 75,000 MCMC samples and discard the first 10,000 samples as burn-in. Using this procedure, we select $M=8$ clusters.

\begin{figure*}
        \centering
        \includegraphics[width=\textwidth]{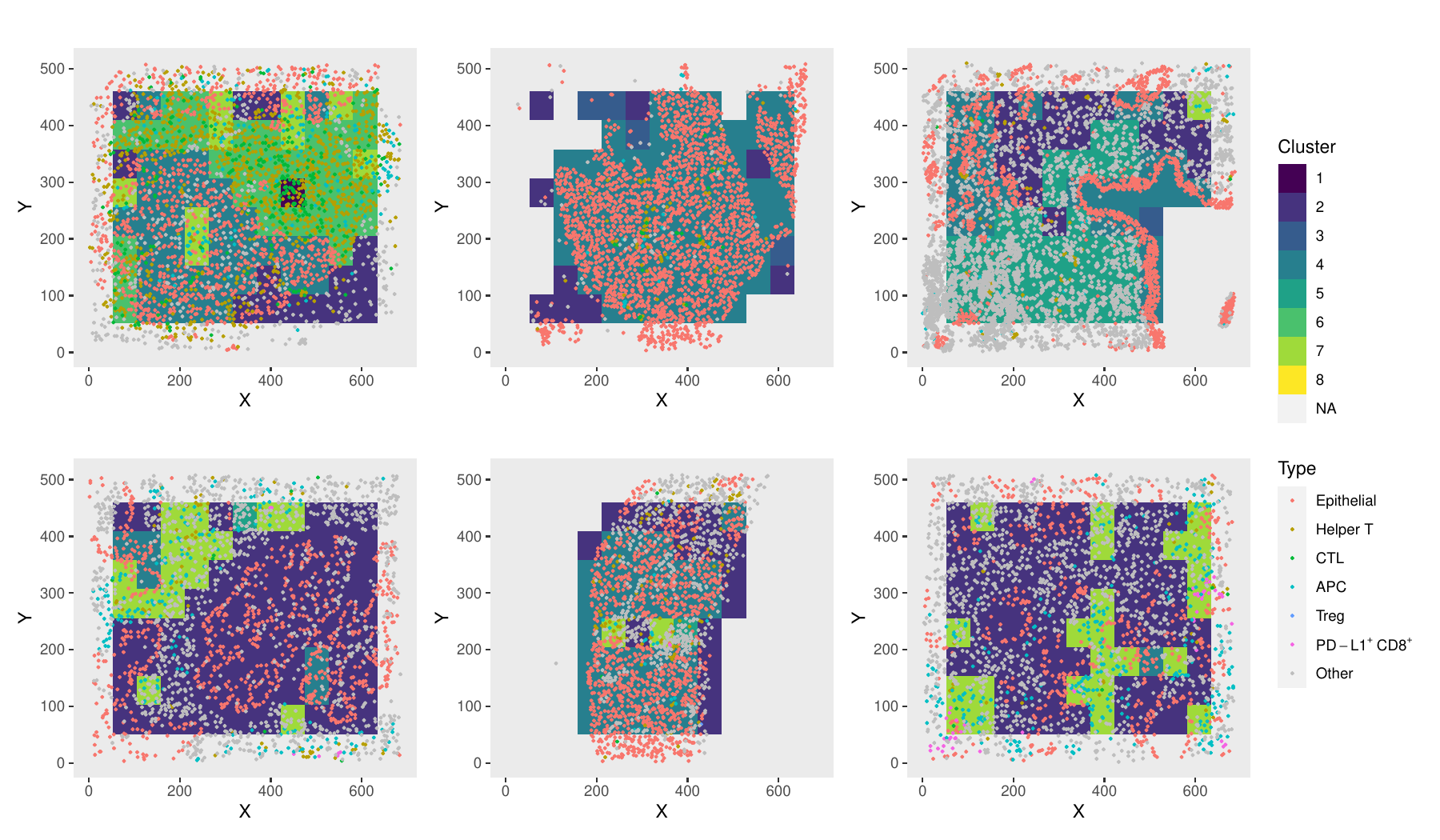}
        \caption{Six representative SPPs obtained from mIF images of pancreatic tissue, overlaid on the estimated cluster labels from the PCM. Points representing cell centroids are colored according to six cell types, as well as other (see legend). The top row shows three SPPs from the non-cancerous disease group and the bottom row shows three SPPs from the cancerous disease group.}
        \label{fig:several_subj_spp_cluster_overlay} 
\end{figure*}

With the fitted PCM, we can segment the SPPs into different clusters and make inferences on the spatial properties of each cluster. 
Figure \ref{fig:several_subj_spp_cluster_overlay} shows six representative SPPs from the mIF dataset overlaid on the estimated cluster labels. There are grid regions with missing data along the boundary of most SPPs, as well as in parts of the SPPs with no cells. However, the PCM assigns cluster labels to these grid regions based on the membership of their neighbors. 

There are significant differences in cluster proportions between cancer groups (Web Table 2). In the non-cancerous disease group, most of the grid regions are split between the clusters 2 and 4, while others are assigned the clusters 5 and 7. In the cancerous disease group, most of the grid regions are assigned cluster 2, while others are assigned the clusters 1, 4, 5, 6, and 7. Thus, the non-cancerous disease group is characterized by two main clusters and two smaller ones, whereas the cancerous disease group is characterized by one main cluster and five smaller ones.

\begin{figure}
\centering
\begin{subfigure}{.5\textwidth}
  \centering
  \includegraphics[width=1\linewidth]{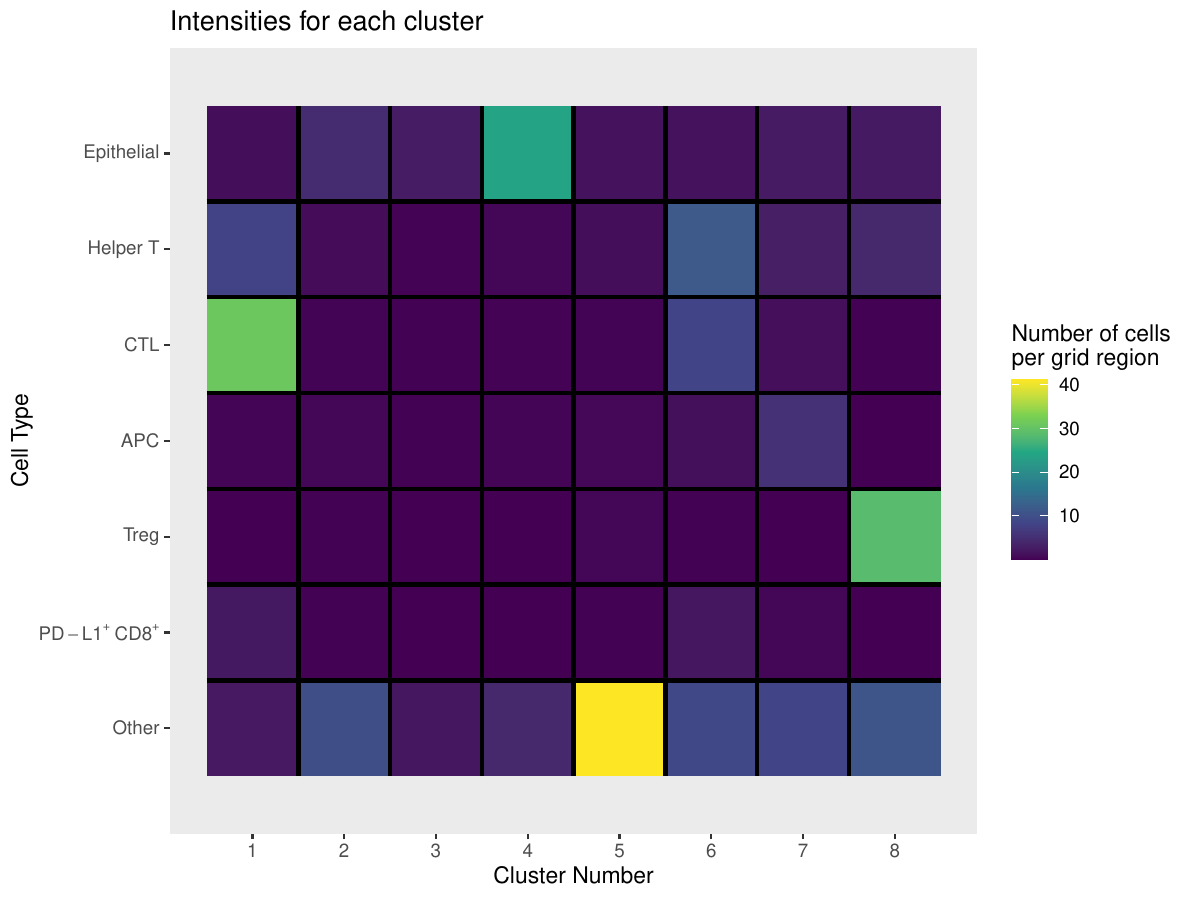}
  \caption{} \label{fig:intensity_summary}
\end{subfigure} 
\begin{subfigure}{.5\textwidth}
  \centering
  \includegraphics[width=1\linewidth]{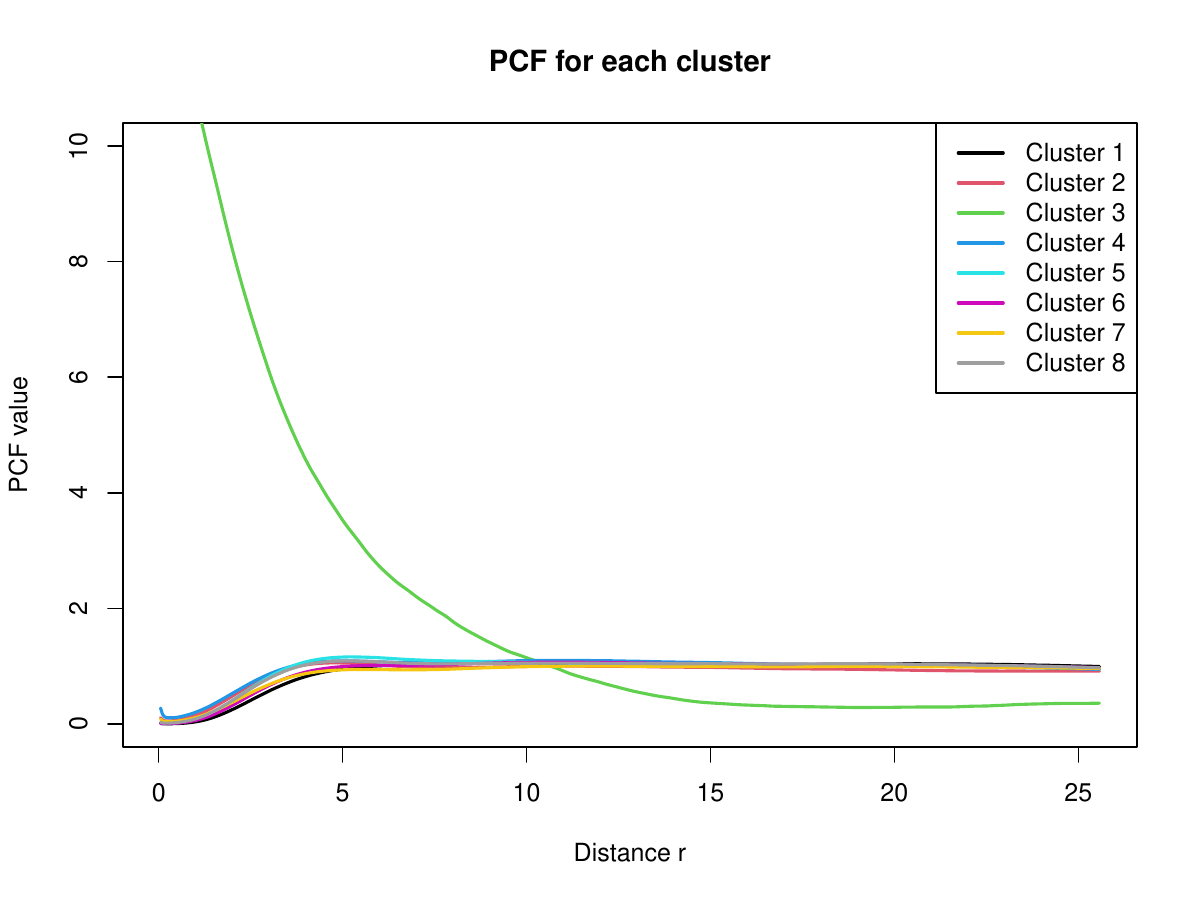} 
  \caption{} \label{fig:PCF_summary}
\end{subfigure} 
\caption{(a) Posterior mean intensities for the seven cell types for each cluster. (b) Posterior mean PCF for each cluster. 
The horizontal red line at 1 represents the theoretical PCF for complete spatial randomness.}
\label{fig:intens_pcf_summary} 
\end{figure}

Figure \ref{fig:intensity_summary} shows the intensity estimates corresponding to the clusters. Cluster 2, which is the main cluster for both groups, is characterized by low intensities for all cell types. However, cluster 4, which is the main cluster for the non-cancerous disease group and a small cluster for the cancerous disease group, is characterized by a relatively high intensity of epithelial cells. The other small clusters (1, 5, 6, and 7) in either group are characterized by relatively high intensities for particular cell types, i.e., CTL, other, helper T, and APC. Therefore, the non-cancerous disease group is characterized by a high intensity of epithelial cells, whereas the cancerous disease group is characterized by high intensities for a more diverse set of cell types. 

\begin{figure*}
    \centering   %
\setkeys{Gin}{width=\linewidth}
\begin{subfigure}{.5\textwidth}
  \includegraphics[width=\linewidth]{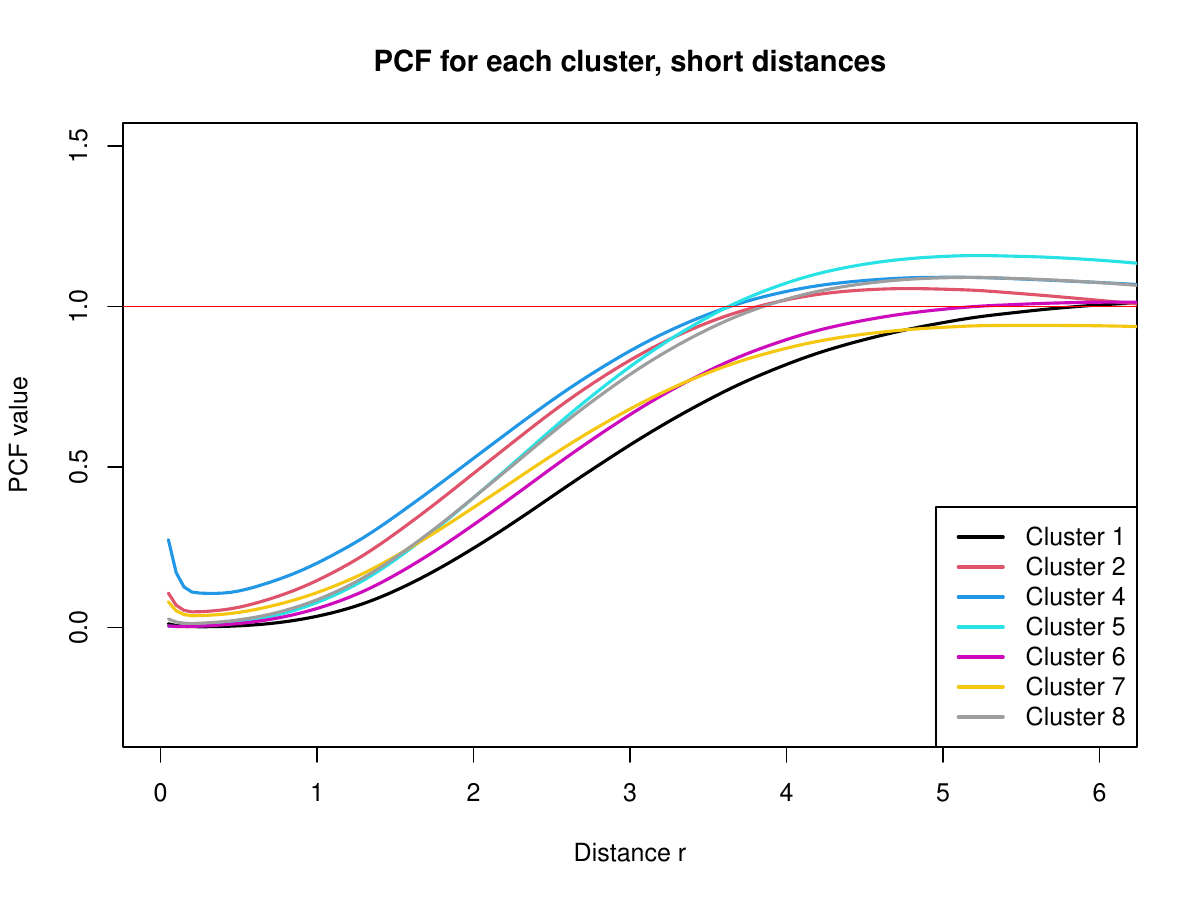}
    \caption{} \label{fig:PCF_summary_restricta}
\end{subfigure}
\hfil
\begin{subfigure}{.44\textwidth}
  \includegraphics[width=\linewidth]{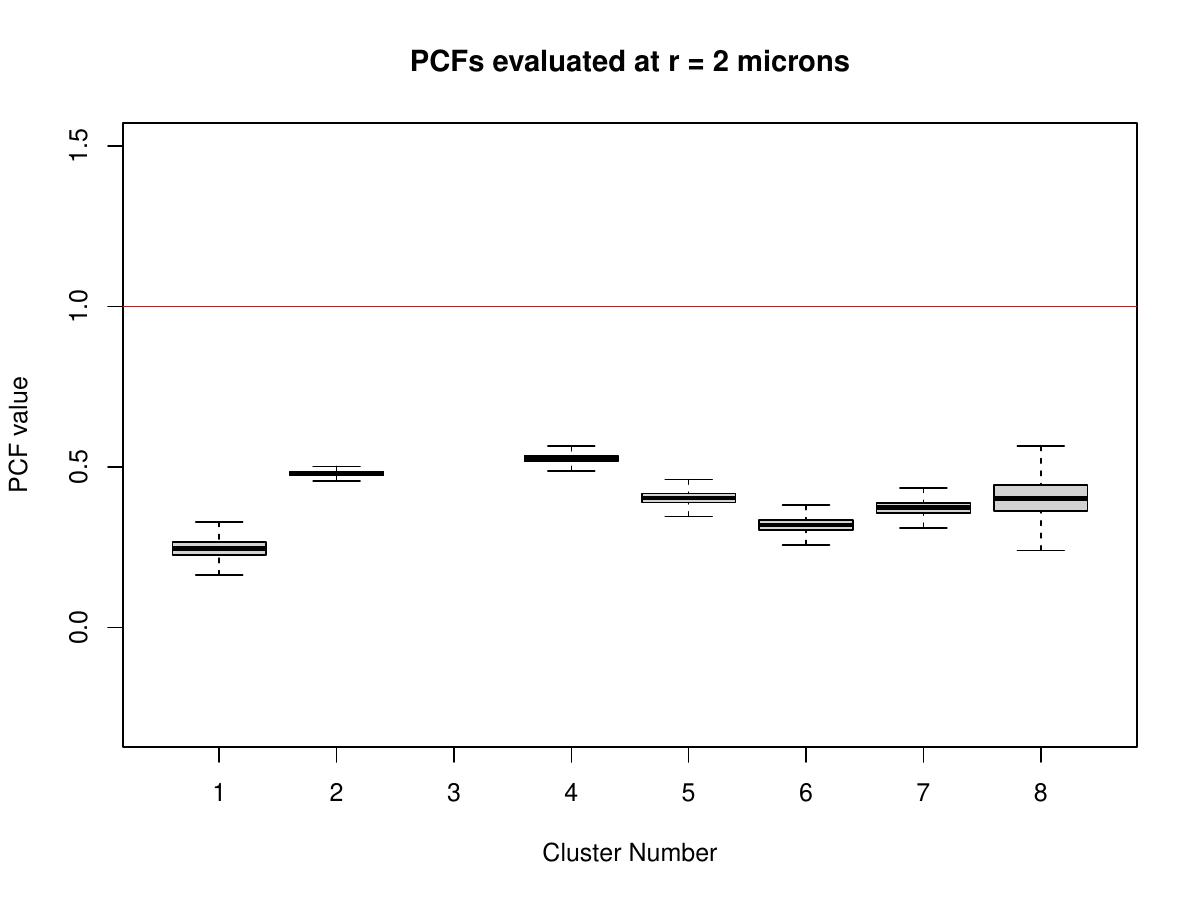}
    \caption{} \label{fig:PCF_summary_restrictb}
\end{subfigure}
\hfil
\begin{subfigure}{.44\textwidth}
  \includegraphics[width=\linewidth]{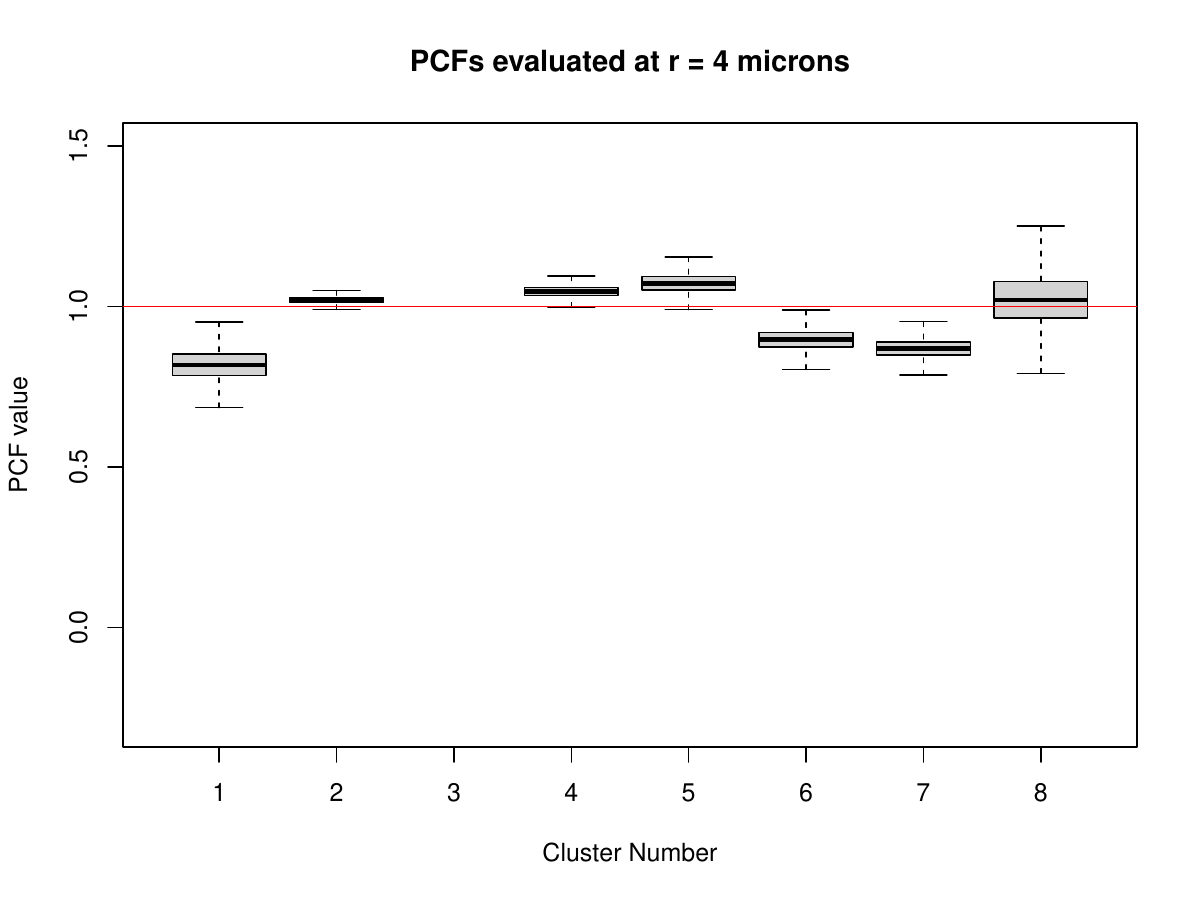}
    \caption{} \label{fig:PCF_summary_restrictc}
\end{subfigure}%
\caption{The PCFs restricted to certain values. The PCF for cluster 3 is omitted in these figures because it is an outlier. The horizontal red line at 1 represents the theoretical PCF for complete spatial randomness. (a) PCFs restricted to the domain 0 to 6 microns. (b) Posterior distribution of PCFs evaluated at 2 microns. (c) Posterior distribution of PCFs evaluated at 4 microns.} \label{fig:PCF_summary_restrict} 
\end{figure*}

Figure \ref{fig:PCF_summary} shows the PCF estimates corresponding to the clusters. Relative to complete spatial randomness, only cluster 3 has a PCF displaying attraction between cells for shorter distances (less than 10 microns), whereas the others display repulsion between cells for shorter distances. However, cluster 3 occurs rarely for both groups. Therefore, we restrict our attention to the PCFs for the other clusters (Figure \ref{fig:PCF_summary_restrict}). As shown in Figure \ref{fig:PCF_summary_restricta}, the PCFs for the clusters display differing degrees of repulsion between cells, relative to complete spatial randomness. We also examine the posterior distributions of the PCFs evaluated at two distances, 2 microns (Figure \ref{fig:PCF_summary_restrictb}) and 4 microns (Figure \ref{fig:PCF_summary_restrictc}). As shown by these figures for either distance, the main clusters 2 and 4 exhibit the least repulsion among cells, whereas the small clusters 1, 6, and 7 have significantly more repulsion between cells. Therefore, the non-cancerous disease group is characterized by less repulsion among cells, whereas the cancerous disease group is characterized by more repulsion among cells.

\section{Discussion}
\label{sec:discuss}

There is need of an interpretable measure of the mIF images' spatial characteristics that can accommodate the size and complexity of such data to drive discovery from the images (see \citet{masotti_dimple_2023}). To this end, we propose a two-stage approach which first extracts estimates of the spatial characteristics of each image according to a grid, then clusters these estimates with the PCM. We test this model with a simulation study and sensitivity analyses, and compare it with several alternatives like $k$-means. We then apply our method to a set of 105 mIF images of diseased pancreatic tissue. 

The application of our method to the mIF images, described in Section \ref{sec:real_data}, provides insight into the tumor microenvironment's spatial properties which can guide further exploratory data analysis. For example, one can explore how the higher intensities of different cell types, apparent in the cancerous disease group, are associated with prognosis. A similar thing is done in the literature, where \citet{nearchou_spatial_2020} developed the spatial immuno-oncology index consisting of the mean CD3\textsuperscript{+} intensity, mean number of lymphocytes in proximity to tumor buds, and ratio of CD68\textsuperscript{+}/CD163\textsuperscript{+} macrophages, and used it as a predictor of prognosis. As another example, one can explore how the varying degrees of repulsion as displayed in Figure \ref{fig:PCF_summary_restrict} are associated with patient outcomes. \citet{Lazarus:2018wu} takes a similar approach where they determine how the interaction between tumor and immune cells can be predictive of survival.

One immediate limitation of our method is the scaling of the computation time of the PCM with the number of clusters. Fitting the PCM with $M>8$ may be infeasible for datasets with 50 subjects or more, as shown in the sensitivity analysis in Web Appendix D.2. In particular, for the motivating dataset with 105 images, running the PCM for nine clusters took five days. Several ways exist to shorten the running time, such as i) fixing the smoothness parameter $\psi$ at the maximum likelihood value in an Empirical Bayes approach, ii) reducing the number of eigenvectors $K$ used to summarize the PCFs, or iii) running multiple shorter MCMC chains in parallel instead of a single, long MCMC chain. 


The proposed approach requires that the resolution of the grid used and the number of clusters are fixed a priori. The resolution of the grid could be underestimated (i.e., made coarser) without a significant decrease in performance, as shown in the sensitivity analysis in Web Appendix D.1. Additionally, if the number of clusters one chooses is slightly above the correct number of clusters, the performance does not change significantly. If running time is not a concern, one can implement Reversible Jump MCMC to estimate the resolution and/or number of clusters within the MCMC \citep{green_reversible_1995}. 

There are a couple of avenues of future work. Our two-stage approach was designed for SPPs from mIF images, but our approach can be straightforwardly extended to SPPs from other kinds of tissue images. For example, \citet{hist_tissue_spatial} show how SPP analysis methods can be applied to H\&E sections. Our method can also be extended to other problems involving a set of SPPs. For example, it can be applied to SPPs consisting of plant locations with different botanical classifications (e.g., see the Lansing Woods dataset \citep{gerrard1969new}), in which each SPP corresponds to a different acre. For applications like these, there may be fewer point types than in tissue images. Thus, we could consider measures of other second-order or higher spatial characteristics, like the pair correlation function measuring the interactions between points arising from different pairs of types (two-type interactions) or even triples of types, which may not be as feasible with many cell types. 

Another avenue of future work would be to posit a Bayesian hierarchical model on the PCFs directly, rather than preprocessing them via principal component analysis. This has the advantage of quantifying the uncertainty in the entire PCF estimate, rather than a low-dimensional representation of it. However, the Bayesian estimation of orthogonal principal components is very challenging; an alternative is to model the PCFs with a spline basis. There are convenient implementations for the Bayesian analysis of penalized splines, e.g., an implementation in WinBUGS \citep{crainiceanu_bayesian_2005}.



\footnotesize

\section*{Supplementary materials}

Web Appendices, Tables, and Figures referenced in Sections \ref{sec:model_framework}, \ref{sec:estimation}, \ref{sec:sim_sensitan}, \ref{sec:real_data}, and \ref{sec:discuss} are available with this paper on arXiv.org. R code and simulated data for one scenario referenced in Section \ref{sec:sim_sensitan} are available for download at https://github.com/AlvinSheng/two-stage-PCM.


\section*{Funding}

This work was supported by a Cancer Center Support Grant Bioinformatics Shared Resource, a gift from Agilent technologies [5 P30 CA046592 to A. R. and S. K.]; a Precision Health Investigator award from University of Michigan Precision Health [to A. R.]; the National Cancer Institute [R37-CA214955 to S. K and A. R.]; University of Michigan startup institutional research funds [to S. K. and A. R.]; and a Research Scholar Grant from the American Cancer Society [RSG-16-005-01 to S. K. and A. R.].

\section*{Conflict of interest}

A. R. serves as a member for Voxel Analytics LLC and consults for Genophyll LLC, Tempus Inc. and TCS Ltd. The remaining authors have no conflicts of interest to disclose.

\section*{Data availability statement}

The dataset of mIF images of diseased pancreatic tissue used in this paper is available from the University of Michigan School of Medicine but is not publicly accessible due to restrictions on licensing for use by the institution. The dataset may be available from the corresponding author on reasonable request, with appropriate permissions from the University of Michigan School of Medicine. 

\normalsize
\bibliography{cip,cell_image_clustering_bib}


\end{document}